\documentclass[twocolumn]{aastex7}

\newcommand{\edit}{}

\usepackage{amsmath}

\begin{document}

\title{Warm Jupiter Tidal Migration Can Spare Inner Planets; Hot Jupiter Tidal Migration May Not}
\author[0000-0002-7733-4522,gname=Juliette,sname=Becker]{Juliette Becker}
\affiliation{Department of Astronomy, University of Wisconsin--Madison, 475 N Charter St, Madison, WI 53706, USA}
\email{juliette.becker@wisc.edu}

\begin{abstract}
In this work, we investigate the dynamical survival of short-period inner planets during the high-eccentricity tidal migration of companion exterior giant planets. Using a combination of analytic arguments and N-body simulations including equilibrium tides and general relativistic precession, we find the boundary in parameter space where an inner companion can remain dynamically stable. We find that survival requires a periastron separation exceeding roughly 14 mutual Hill radii at closest approach. Below this threshold, secular eccentricity exchange, orbit crossing, and/or tidal evolution can lead to the destruction of the inner planet. We apply our methodology to the current exoplanet sample and find that none of the known systems containing a short-period giant and an inner companion could have assembled via high-eccentricity tidal migration. However, warm Jupiters with larger periastron distances ($q_{\mathrm{out}} \sim 0.05-0.08$ AU, corresponding to final observed semi-major axis values $a_{\mathrm{out}} \sim 0.10-0.16$ AU) can allow the survival of short-period inner planets while potentially also circularizing on $\lesssim 1$~Gyr timescales. Our results provide a framework for distinguishing disk migration from tidal migration in observed multi-planet systems containing close-in gas giants.
\end{abstract}



\section{Introduction} 
Hot Jupiters are gas giant planets on short-period orbits, and are widely believed to reach these orbits primarily through high-eccentricity tidal migration, a mechanism thought to account for an estimated 60–84\% of observed hot Jupiters \citep{Petrovich2016, Zink2023}. 
In this scenario, a Jupiter-mass planet first forms beyond the ice line of its host star and later undergoes a significant increase in orbital eccentricity.
This eccentricity excitation could occur due to 
interactions with a companion planet \citep{Nagasawa2008, Beauge2012}, \edit{a bound companion star \citep{Fabrycky2007, Anderson2016, Noaz2016}}, a stellar fly-by \citep{Hamers2017, Rodet2021}, or by more complex mechanisms such as eccentricity cascades driven by multi-body dynamics \citep{Yang2025}.
Once the Jupiter's eccentricity becomes sufficiently high, its orbit gradually circularizes due to tidal interactions with its host star \citep[e.g.,][]{Hut1981}.

However, the existence of nearby planetary companions to some hot Jupiters \citep[e.g.,][]{Becker2015, Canas2019,Huang2020, Hord2022, Sha2023, Maciejewski2023, Korth_2024} poses a potential challenge to this scenario. A tidally migrating Jupiter may reach eccentricities in excess of $e\sim0.9$ \citep{Wu2003}.  
The presence of such companions is often interpreted as evidence against high-eccentricity migration, since this process is typically thought to be dynamically disruptive and unlikely to preserve close-in neighbors. 
As a result, for outer companions to hot Jupiters, there is very little parameter space in a planetary system where a companion might avoid orbit crossing during the migration process. Inner companions, as addressed in \citet{Mustill2015}, are usually similarly endangered: the planetary orbital radii typical in the transiting compact multi-planet sample, $0.05~-~0.5$ AU \citep{Fang2012, Howe2025}, reside near or exterior to the final orbital location of the hot Jupiter. 
Consequently, systems with both a hot Jupiter and a nearby planetary companion are usually assumed to have formed through disk-driven migration, which avoids orbit crossing and is comparatively gentle and can maintain initially coplanar, multi-planet architectures.

There are two ways to avoid instabilities from true or near orbit crossing:
a) If the migrating Jupiter stops at a slightly larger orbital distance (a ``warm" Jupiter), it may avoid physically crossing the orbits of the inner planets.
b) If the inner planet has an extremely short orbital period, it can become dynamically decoupled \edit{(due to tides, GR, or stellar gravitational harmonics)} and survive despite the Jupiter’s migration.
In this work, we consider these possibilities by exploring whether systems with hot (or warm) Jupiters and extremely short period inner companions could still be the product of high-eccentricity tidal migration. 
Specifically, we pose the following question: if a system begins with an extremely short-period inner planet ($a\sim 0.01-0.03$ AU) and a distant giant planet, and the outer giant later undergoes high-eccentricity tidal migration, under what conditions can the inner planet survive this process? 

In Section \ref{sec:analytic}, we provide some motivating analytic estimates of when inner companions might survive this process, and apply these analytic estimates to the exoplanet sample. In Section \ref{sec:numeric}, we present the results of a suite of numerical simulations that provide boundaries on the parameter space where inner planets may survive. In Section \ref{sec:discuss}, we consider our results in the context of the exoplanet sample. We conclude in Section \ref{sec:conclude} with a summary of our results.

\begin{figure}[ht!]
\includegraphics[width=0.42\textwidth]{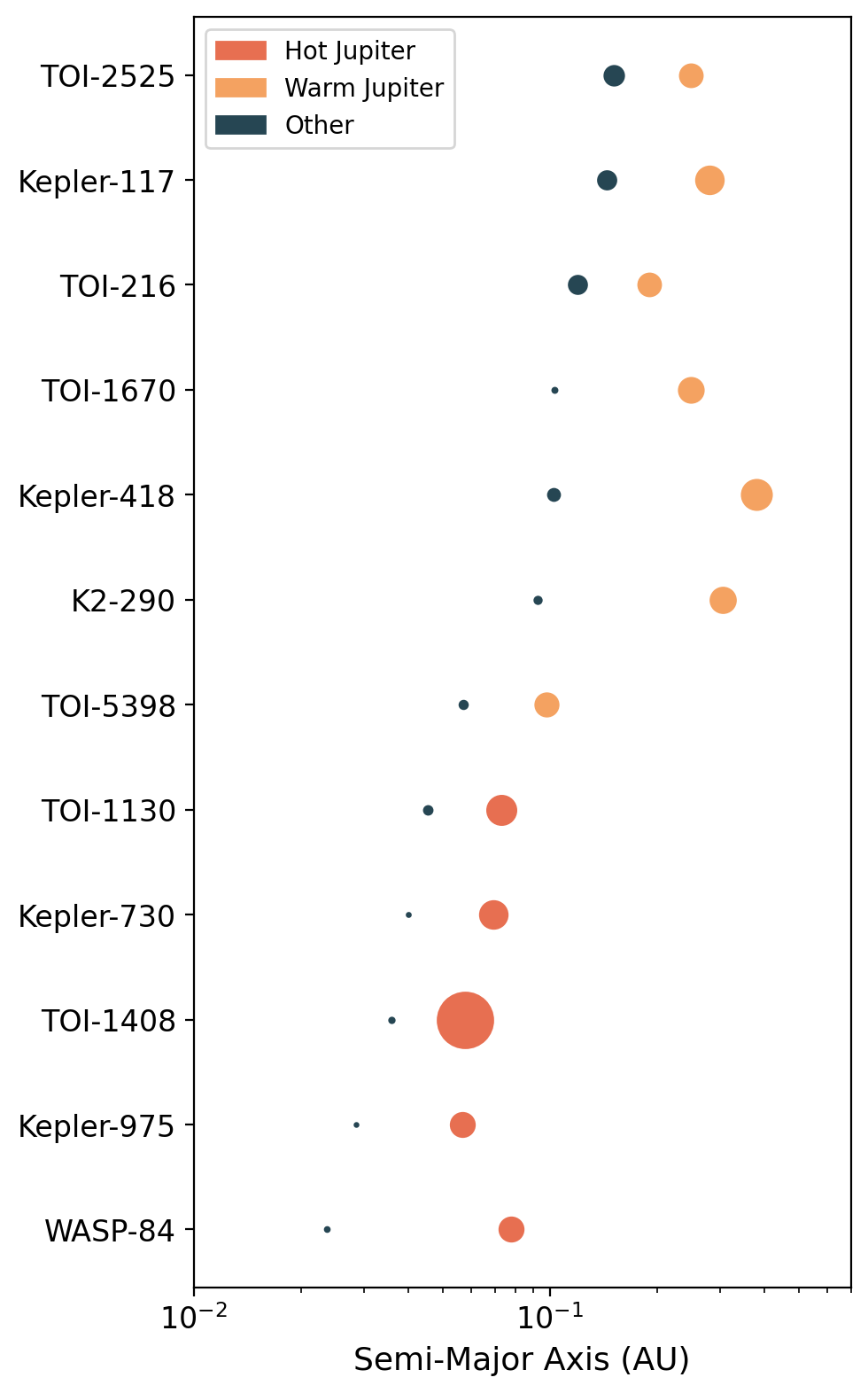}
    \caption{Schematic representation of planetary systems known to contain an inner small planet and an outer short-period giant planet (hot or warm Jupiters) and no other nearby planets. The horizontal axis shows the semi-major axis of each planet. Marker size scales with planet radius. Hot Jupiters (orange) and warm Jupiters (yellow) are distinguished from smaller inner companions (dark blue). Data obtained from the NASA Exoplanet Archive \citep{Christiansen2025}, 05/13/2025.}
    \label{fig:schematic}
\end{figure}

\begin{figure}[ht!]
\includegraphics[width=0.48\textwidth]{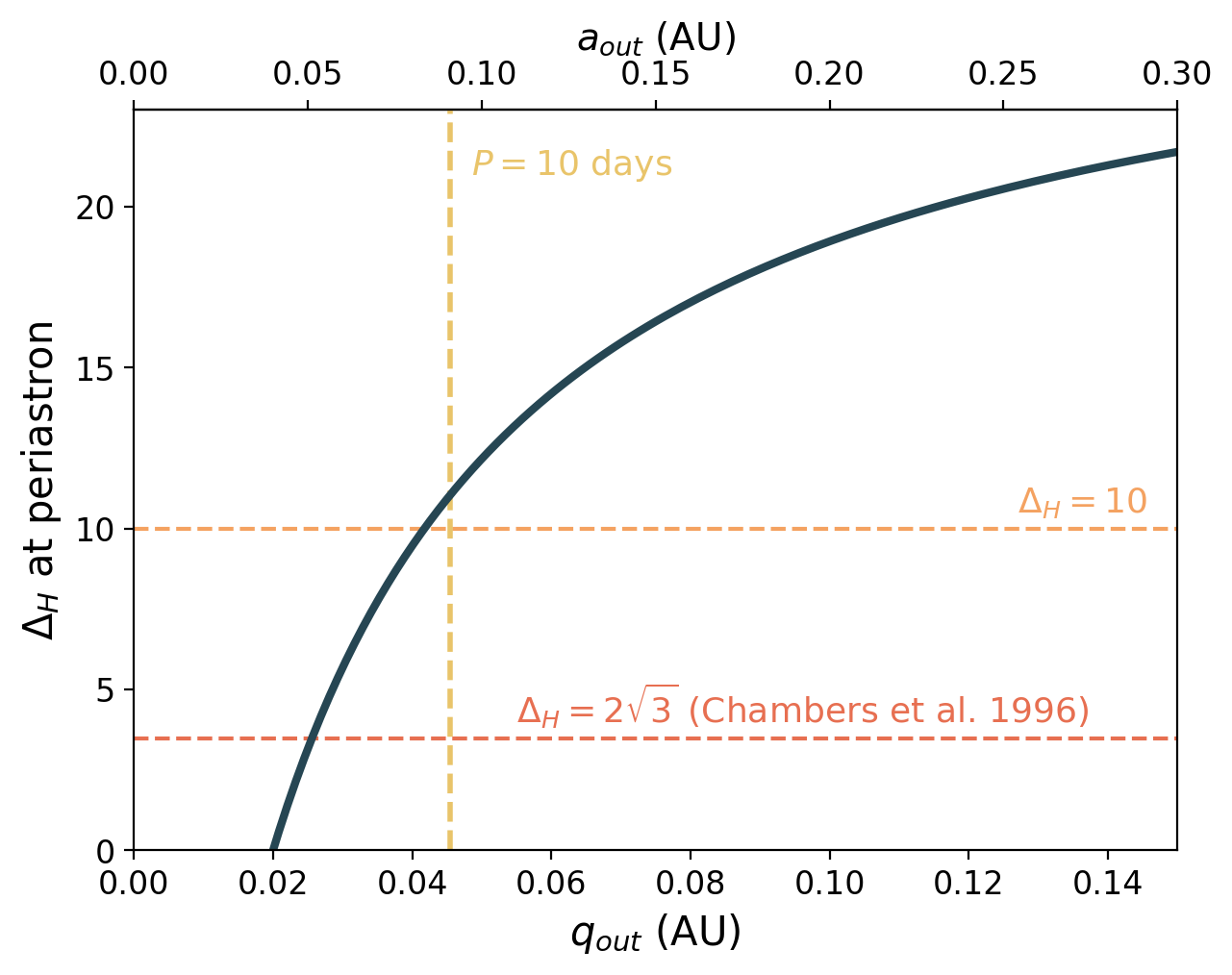}
\caption{Mutual Hill‐radius separation $\Delta_H$ at periastron between a fixed inner planet ($a_{\rm{in}} = 0.02\,$AU, $m_{\rm{in}}=1\ M_\oplus$) and a migrating outer Jupiter-mass planet ($m_{\rm{out}}=1 \ M_{J}$, initial $a_{\rm{out}} = 2$ AU) around a $1\,M_\odot$ star.  The bottom x-axis shows the outer Jupiter’s periastron distance $q_{\rm{out}}$ during tidal migration, while the top x‐axis gives the corresponding final semi-major axis $a_{\rm{out}}$ after tidal migration is complete.  The horizontal dashed lines mark common literature stability thresholds at $\Delta_H = 2\sqrt3$ \citep{Chambers1996} and $\Delta_H=10$. The vertical dashed line indicates the periastron corresponding to a 10 day orbital period, the commonly used demarcation between hot and warm Jupiters.  }
\label{fig:hill_radius_theory}
\end{figure}

\begin{figure*}[ht!]
\centering
\includegraphics[width=0.9\textwidth]{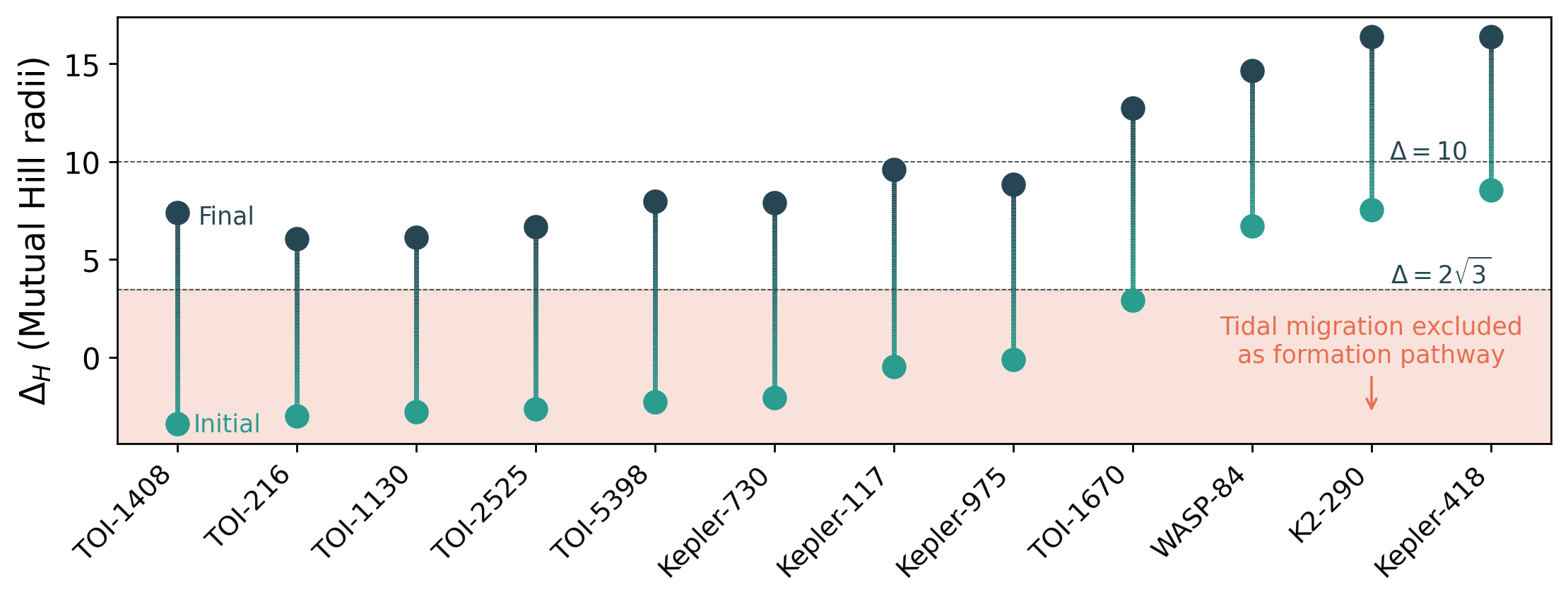}
\caption{Mutual Hill–radius separations $\Delta_H$ for each two‐planet system, sorted by their initial separation $\Delta_{H,i}$ along the horizontal axis. \edit{Teal} circles mark the initial $\Delta_{H}$; those falling below zero (shaded red region) represent systems where the orbits would have crossed during tidal migration, thus excluding that assembly pathway. \edit{Black} circles show the current-day observed separations $\Delta_{H}$, and vertical lines connect each system’s initial and final values. Dashed horizontal lines at $\Delta=2\sqrt{3}$ and $\Delta=10$ denote two typical literature stability thresholds in units of mutual Hill radii. Planet parameters obtained from \citet{Tingley2014, Bruno2015,Morton2016,  Hjorth2019, Huang2020, Canas2019, Dawson2021, Valizadegan2022, Tran2022, Trifonov2023,Maciejewski2023, Mantovan2024, Korth_2024, Borsato2024}.}
\label{fig:hill_spacing}
\end{figure*}

\section{Analytic Motivation}
\label{sec:analytic}
When a hot Jupiter undergoes inward tidal migration, it can destabilize nearby companion planets through two primary mechanisms. In systems with either interior and/or exterior companions, instability may arise due to direct orbit crossing. For some subset of interior companions, however, the periastron distance of the migrating Jupiter may not physically cross the orbit of the inner planet. The migrating planet could in this case instead induce strong planet–planet interactions. These interactions allow for the exchange of orbital eccentricity between planets \citep{MD99}, which can lead to collisions or drive the inner planet onto a high-eccentricity orbit that brings it perilously close to the star \citep{Carrera2019}, in which case tidal forces may then cause the inner planet to spiral into the star and be destroyed \citep[e.g.,][]{Hamer2019}.

However, such a dynamical instability is not inevitable. If the periastron of the outer migrating giant planet remains well-separated from the apastron of the inner companion, the two planets can avoid destabilizing close encounters, potentially allowing the inner planet to survive the migration of its more massive neighbor.

To estimate the parameter space where this may occur, we can first compute the final semi-major axis of a tidally migrating planet as a function of its initial periastron distance, under the assumption that angular momentum is conserved and energy is dissipated via heat in the planetary envelope \edit{or in the stellar envelope}.
We assume that the mass of the migrating planet remains constant during migration. In that case, we can write its specific orbital angular momentum as
\begin{equation}
h = \sqrt{G M a (1 - e^2)},
\end{equation}
where $G$ is the gravitational constant, $M$ the mass of the central body, $a$ the planet's semi-major axis, and $e$ its orbital eccentricity. 
Conservation of angular momentum during the migration process requires
\begin{equation}
\sqrt{G M a_i (1 - e_i^2)} = \sqrt{G M a_{f}}.
\end{equation}
where subscripts $i$ and $f$ denote the initial and final values of the planetary orbit. Solving this expression for the final orbital radius, $a_f$, yields
\begin{equation}
a_f = a_i(1 + e_i)(1 - e_i) = q_i(1 + e_i),
\end{equation}
where $q = a (1-e)$ is the periastron distance of the planet. 
We assume that the initial scattering event driving the onset of tidal migration excited the eccentricity of the Jupiter-mass planet to large values. 
Under that assumption, we can take the limit where $ e_i \to 1$, in which case
\begin{equation}
\lim_{e_i \to 1} a_f = \lim_{e_i \to 1} q_i (1 + e_i) = 2 q_i
\label{eq:ang_conservation}
\end{equation}

This estimate, which connects the initial periastron distance of a migrating Jupiter to its final semi-major axis \edit{under the assumption that angular momentum of the planet's orbit is conserved}, allows us to infer the closest past approach ($q_i$) between an observed Jupiter-mass planet and an interior companion using only their currently measured orbital parameters ($a_i$). \edit{It is important to note that in some systems, particularly those with shorter periastron distances or larger planet masses, some angular momentum from the planetary orbit will go into decreasing the spin period of the star, resulting in slightly larger final semi-major axis values than predicted by Eq. \ref{eq:ang_conservation}. }

To determine whether significant interactions likely occurred during this closest approach, we can compare their inferred closest past separation to the mutual Hill radius.
The mutual hill radius $R_{H,m}$ is the characteristic distance at which the gravitational spheres of influence of two adjacent planets overlap \citep{Gladman1993, Chambers1996, Marzari2002, Morrison2016}, and it is written as
\begin{equation}
R_{H,m} = \left(\frac{\mu_{i-1} + \mu_{i}}{3}\right)^{1/3} \,\frac{a_{i-1} + a_{i}}{2},
\label{eq:basicRH}
\end{equation}
where $\mu_i = m_{i} / M$ is the mass ratio between planet $i$ and the central body, and $a_{i-1}$ denotes the semi-major axis of the interior companion to the migrating Jupiter-mass planet.   
When planets reside closer to each other in units of this mutual hill radius, they are likely to interact \citep{Gladman1993} and very closely spaced orbits will likely go dynamically unstable over long time scales \citep{Obertas2017}. 

This work focuses on systems consisting of two known planets with an outer planet of roughly Jupiter mass. Figure \ref{fig:schematic} shows a diagram of known systems with such geometries. 
For the situation we consider in this work, where an outer planet in a two-planet system has a significant orbital eccentricity, we can adapt Equation \ref{eq:basicRH} to account for the close approach, which occurs when the outer planet reaches periastron $q_{\rm{out}} = a_{\rm{out}}(1-e_{\rm{out}})$: 
\begin{equation}
\Delta_{H}^{\rm peri}
= \left(\frac{\mu_{\text{in}} + \mu_{out}}{3}\right)^{\!1/3}
\,\frac{a_{\rm{in}} + a_{\rm{out}}(1 - e_{\rm{out}})}{2}\,.
\label{eq:deltaH}
\end{equation}
In Figure \ref{fig:hill_radius_theory}, we show how for a idealized two-planet system how the spacing between the planets (measured in units of $\Delta_{H}^{\rm peri}$) changes with the distance of closest approach of the outer migrating Jupiter. We note that the orbital period $p \approx 10$ day dividing line between hot and warm Jupiters, which was determined informally by the astronomical literature, happens to be very close to the $\Delta_H^{peri} = 10$ line for an inner companion at $a_{\rm{in}} =0.02$ AU.

Then, in Figure \ref{fig:hill_spacing}, we show the mutual Hill radius separation between each pair of planets shown in Figure \ref{fig:schematic}, specifically systems consisting of a hot or warm Jupiter with a single interior companion. For each system, we plot both the final observed separation (dark blue point) and the inferred initial mutual Hill radius separation that would have occurred if the Jupiter attained its current orbit via tidal migration.

We shade the region in Figure \ref{fig:hill_spacing} where tidal migration is likely excluded as a viable formation pathway due to dynamical considerations. For systems where the initial mutual Hill radius separation is negative, the migrating Jupiter would have crossed the orbit of the inner companion. This implies that orbital crossing would have occurred, making tidal migration impossible for those systems.
Similarly, for systems with $\Delta_H < 2\sqrt{3}$, the threshold identified by \citet{Chambers1996} for long-term dynamical instability, repeated interactions during the Jupiter’s migration would likely have destabilized the inner companion. As such, tidal migration is inconsistent with the orbital architecture of systems where the inferred initial Hill radius separation falls below this threshold.
However, the separation threshold given in \citet{Chambers1996} is the absolute lower limit, below which systems are almost certain to go dynamically unstable. Simulation work over the past decade has identified that in tightly packed exoplanet systems, a lager threshold ($\Delta_H \sim 8-10$) may be required to maintain dynamical stability over long terms \citep{Obertas2017, Lissauer2021}. As such, while the analytic threshold shown in Figure \ref{fig:hill_spacing} is sufficient to identify systems that absolutely could not have assembled via tidal migration (the leftmost nine systems in the figure), it does not follow from this that the remaining three systems (WASP-84, K2-290, Kepler-418) have hot/warm Jupiters that attained their final orbits via tidal migration. We will evaluate the feasibility of that in Section \ref{sec:casestudies}.

\begin{figure}[ht]
\centering
\includegraphics[width=0.45\textwidth]{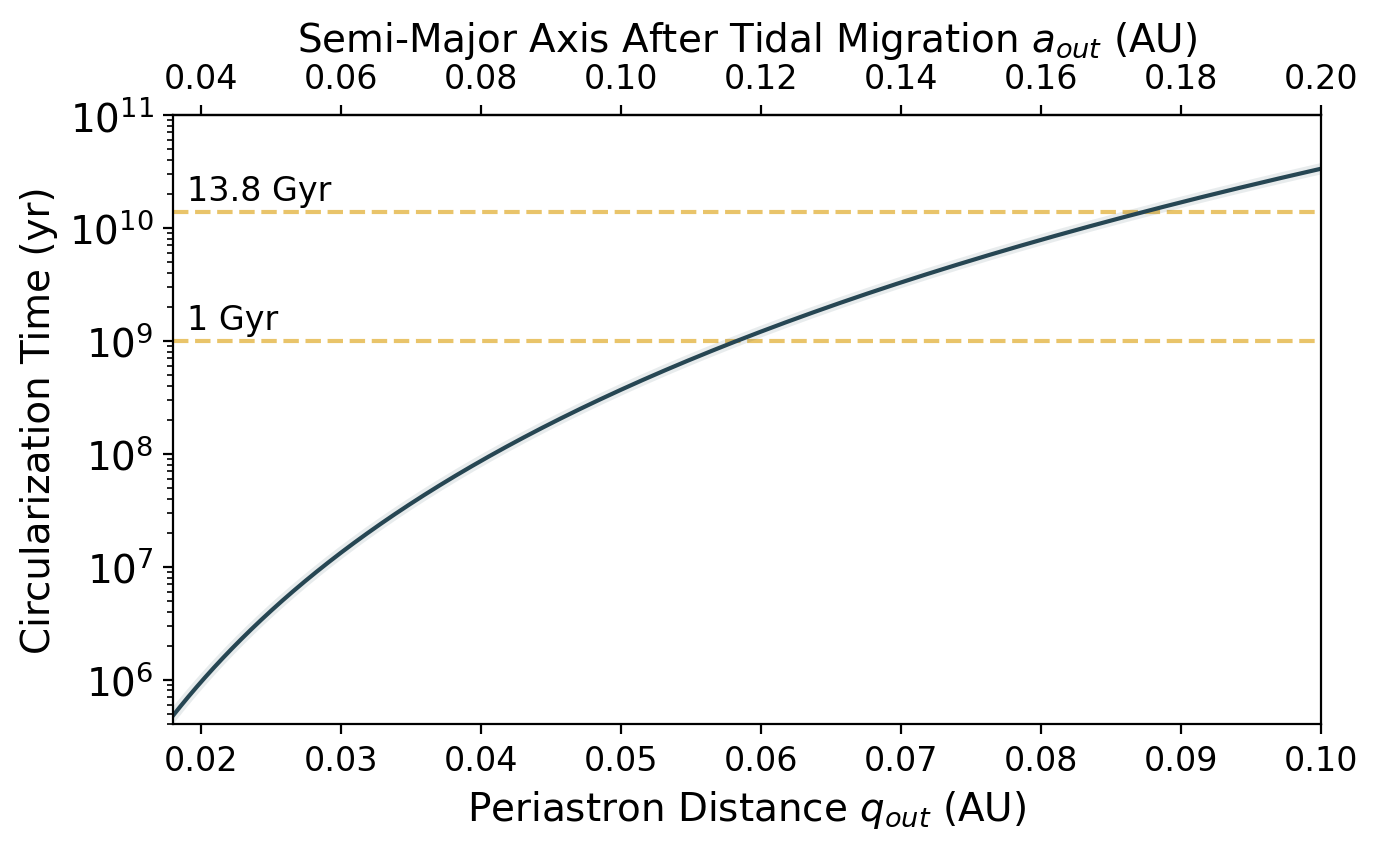}
\caption{
Tidal circularization timescales, computed using Equation \ref{eq:tcirc}, as a function of periastron distance $q_{\mathrm{out}}$ for a highly eccentric ($e = 0.9$) hot Jupiter orbiting a Sun-like star. The planet is assumed to have Jupiter-like mass and radius ($M_p = 1\,M_{jup}$, $R_p = 1\,R_{jup}$), while the host star has mass $M_\star = 1\,M_\odot$ and radius $R_\star = 1\,R_\odot$. The tidal quality factors are set to $Q_p = 10^4$ and $Q_\star = 10^6$, with a planetary Love number $k_2 = 0.5$. The secondary x-axis shows the corresponding final semi-major axis for the hot Jupiter after full circularization according to Equation \ref{eq:ang_conservation}. All timescales are computed assuming non-time-varying planetary and stellar structure.
For these fiducial values, proto-hot Jupiters outside of periastron distance $q_{\mathrm{out}}\sim0.06 - 0.08$ are unlikely to circularize by the time that we observe them. }
\label{fig:circ_time}
\end{figure}

\section{Simulation Verification}
\label{sec:numeric}
The analytic arguments in Section \ref{sec:analytic} suggest that a inner planet may survive the tidal migration of an exterior Jupiter companion if the separation at the periastron of the outer planet is significantly large, measured in units of the mutual hill radius. 
In this section, we first test this hypothesis with numerical simulations and identify the part of parameter space where inner companions may survive. 
We then perform a second set of targeted simulations to test whether the inner companions in the known systems WASP-84, K2-290, and Kepler-418 (whose outer Jupiters could not be ruled out as having undergone tidal migration based on our previous analytic arguments) could have survived such migration.

\subsection{Simulation Setup}
To evaluate the survival of inner companions during the high-eccentricity migration of exterior Jupiter-mass companions, we perform numerical integrations using the \texttt{REBOUND} package \citep{Rein_2012} and the \texttt{REBOUNDx} \citep{Tamayo2020} implementation of the equilibrium tide \citep{Lu2023}.  
The simulations used the \texttt{mercurius} integrator and included the first post-Newtonian (1PN) correction from general relativity (GR) on the central body \citep{gr_potential, Tamayo2020} and the effects of tidal evolution \citep{Baronett2022, Lu2023}. This correction is important for accurately modeling the stability of these strongly interacting but extremely short-period multi-planet systems \citep{Adams2006, Faridani2022}. The simulation timestep for the \texttt{WHFAST} portion of the integrator was set to $1/30$ of the innermost planet's orbital period.

The simulations were initialized with two planets: an inner planet with mass $m_{\rm{in}} = 1 M_{\oplus}$, radius $R_{\rm{in}} = 1 R_{\oplus}$, an initial eccentricity $e_{\rm{in}} = 0$ and a semi-major axis $a_{\rm{in}}$ which is allowed to vary, and an outer planet with mass $m_{\rm{out}} = 1 M_{J}$, radius $R_{out} = 1 R_{J}$, semi-major axis $a_{\rm{out}} = 2$ AU, and an initial eccentricity $e_{\rm{out}} \in [0.95, 0.99]$ which is allowed to vary and sets the initial periastron distance $q_{\rm{out}}$ of the tidally migrating Jupiter.  
\edit{In all integrations, the star was initialized with a solar-like spin period of 27 days, and no obliquity relative to the plane containing the planetary orbits. The planets were initialized with spin periods of 0.5 days, and 0 degree obliquity relative to their orbital planes. }

\begin{figure*}[ht!]
\centering
\includegraphics[width=0.97\textwidth]{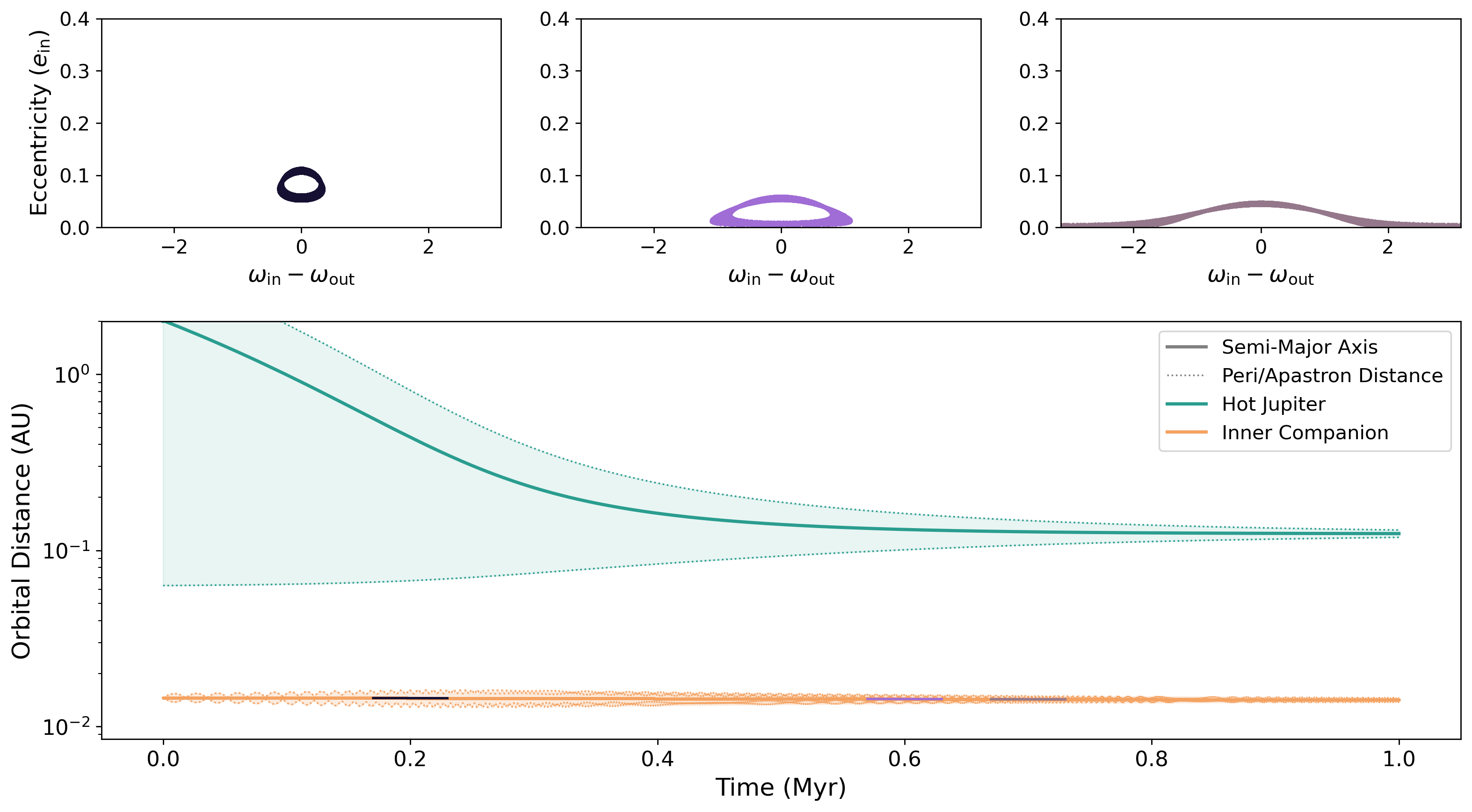}
\caption{One illustrative simulation from our sample, showing a dynamically stable evolution in which an inner companion at $a_{\rm{in}} \approx 0.015$ remains stable, bound, and with relatively low orbital eccentricity as its outer Jupiter-mass companion tidally migrates from an initial periastron distance $q_{\rm{out}} \approx 0.063$ AU. \emph{Top panel:} The evolution of the inner planet in $(\omega_{\mathrm{in}} - \omega_{\mathrm{out}}, e_{\mathrm{in}})$ phase space at three distinct times during the integration, illustrating the transition of the inner planet's dynamics from planet-planet induced libration at early times to circularization driven by stellar effects at late times. \emph{Bottom panel:} The orbital evolution of both planets over time. Solid lines represent the semi-major axes, while the shaded regions indicate the range between periastron and apastron distances. 
}
\label{fig:stability}
\end{figure*}

When testing hot Jupiters with relatively larger periastron distances, the tidal circularization timescales become quite large, as the circularization timescale scales as $\tau_{\mathrm{circ}} \propto a^{-13/2}$ \citep{Goldreich1966, Driscoll2015, Barnes2020}. This presents a computational challenge: our simulations must use short timesteps to accurately resolve the ultra-short-period planet's orbit and close encounters at periastron, while also integrating over very long timescales to capture the full extent of tidal circularization.

To address this, we rescale the circularization timescale by adjusting the tidal quality factor $q_{\rm{out}}$ of the hot Jupiter. We choose the value of $q_{\rm{out}}$ such that the outer planet's circularization time is 1 million years for all integrations (while planet-planet interactions may slightly alter the exact timescale). This results in every integration, where $e_{\rm{out}}$ varies, having a different tidal $q_{\rm{out}}$ factor.
\edit{We note that this approximation, while necessary to make the problem computationally tractable for a large parameter sweep, will potentially underestimate the range of parameter space where stable migration can occur, as longer integrations would allow for slower, potentially more adiabatic orbital evolution.}

The circularization time for planets on eccentric orbits can be written as \citep{Matsumura2008}: 
\begin{equation}
\tau_{\mathrm{circ}}
\;\equiv\;\frac{e}{|\dot e|}
\;\approx\; \frac{1}{27}\,
\frac{Q_p}{n\ k_2}\,
\frac{M_p}{M_\star}\,
\left(\frac{R_p}{a}\right)^{-5}
\,F_p^{-1}\,.
\label{eq:tcirc}
\end{equation}
when
\begin{equation}
\begin{split}
F_p
= \biggl[&
  \frac{1 + \tfrac{15}{4}e^{2} + \tfrac{15}{8}e^{4} + \tfrac{5}{64}e^{6}}     {(1 - e^{2})^{13/2}}\\
  &\;-\;
  \frac{11}{18}\,
  \bigg(\frac{1 + \tfrac{3}{2}e^{2} + \tfrac{1}{8}e^{4}}
     {(1 - e^{2})^{5}}\bigg)
  \frac{\Omega_{*}}{n}
\biggr]\,.
\end{split}
\end{equation}
For the purposes of this calculation, we assume $\Omega_{*} = n$.
To keep our integration times consistent, we choose $q_{\rm{out}}$ individually for each integration such that $\tau_{\mathrm{circ}} = 1$ Myr, as follows: 
\begin{equation}
Q_{\rm{out}} = 27\,\tau_{\mathrm{circ}} \frac{n\ k_2\ M_\star}{m_{\mathrm{out}}}\ \left(\frac{R_{\mathrm{out}}}{a}\right)^{5}
\,F_p\,.
\end{equation}
We set the tidal quality factor of the inner planet to be $Q_{\rm{in}} = 10^{-3}\ Q_{\rm{out}}$, assuming the inner planet is a super-Earth or sub-Neptune and proportionately thus more dissipative. 

\edit{The tidal quality factor \( Q \) is expected to vary as the orbital frequency \( n \) evolves \citep[e.g.,][]{Naoz2012}. However, allowing \( Q \) (or equivalently the time lag \( \tau \)) to evolve causes the circularization timescales to vary non-uniformly between simulations. To maintain numerical consistency and ensure that all systems circularize within the prescribed integration time, we adopt a constant \( Q \). This simplification means that the absolute circularization times are not physically meaningful, but the relative dynamical behavior and stability outcomes remain valid.}
\edit{Similarly, changing the stellar spin period would further alter the circularization time. However, because we rescale all circularization times to a fixed integration length of 1~Myr, this effect does not influence the our simulation outcomes. } 
\edit{In the \texttt{REBOUNDx} implementation of equilibrium tides, uniformly rescaling the tidal quality factors ($Q$) for all bodies leaves the system’s trajectory in orbital/spin phase space unchanged; only the rate at which the system moves along that trajectory is altered.}

\citet{Zink2023} finds that most hot Jupiters migrate via coplanar high eccentricity migration. As a result, we initialize our planets on coplanar orbits. This setup also increases the chances of a mutually transiting geometry, which is necessary for the planets to be found, as systems of the geometry under consideration have (so far) only been discovered via transits. 

Following the basic simulation scheme outlined above, we conduct a parameter sweep over semi-major axis of the inner planet $a_{\rm{in}} \in [0.01, 0.04]$ AU and periastron distance for the migrating Jupiter $q_{\rm{out}} \in [0.02, 0.1]$ AU. We run a total of 450 simulations, with values randomized within this range.

\begin{figure*}[ht!]
\centering
\includegraphics[width=0.97\textwidth]{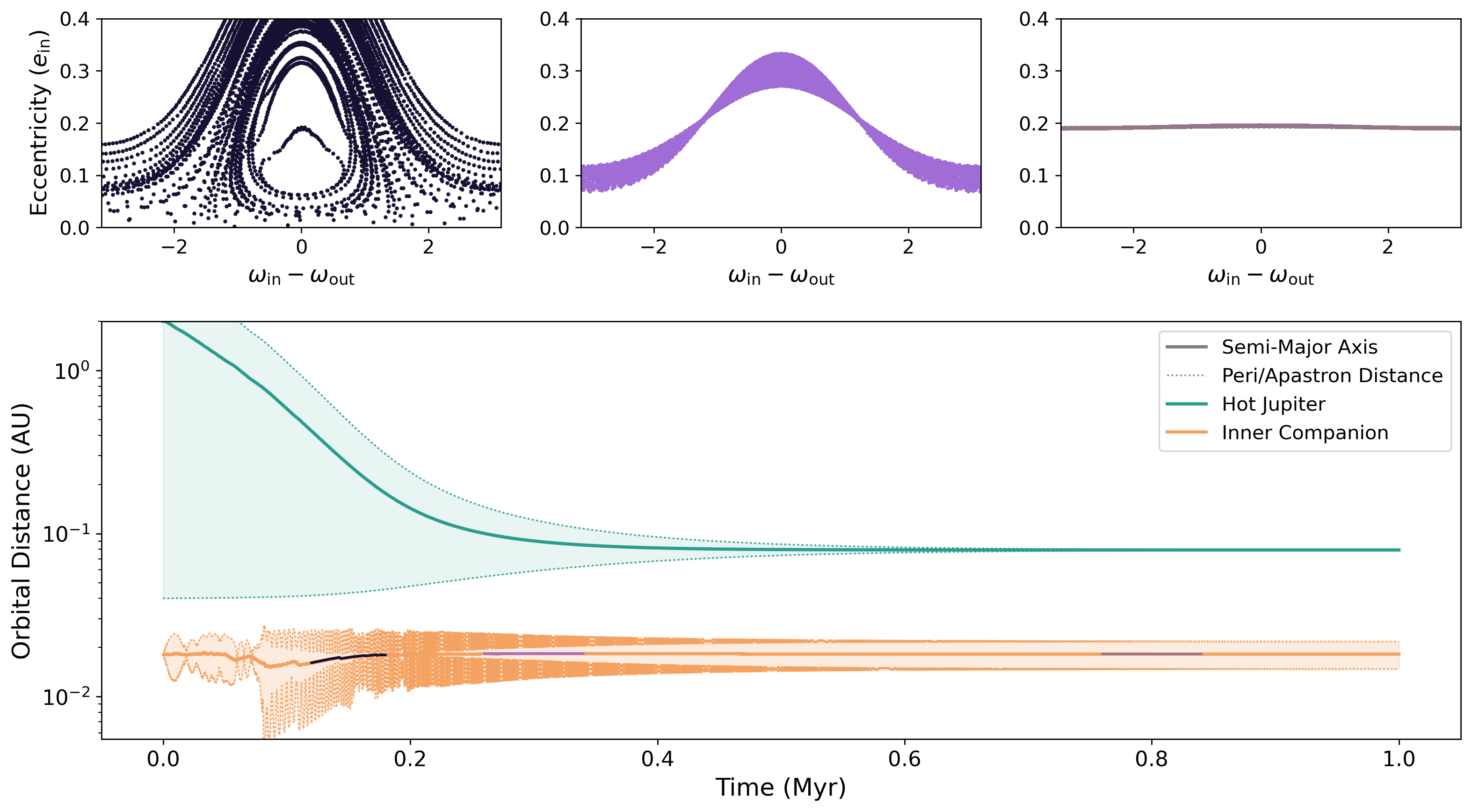}
\caption{A second illustrative simulation from our sample, showing a dynamically stable evolution with stronger interactions than the integration shown in Figure \ref{fig:stability}. This integration contained an inner companion at $a_{\rm{in}} \approx 0.018$ and an outer companion with an initial periastron distance $q_{\rm{out}} = 0.04$. 
\emph{Top panel:} The phase space evolution of the inner planet in $(\omega_{\mathrm{in}} - \omega_{\mathrm{out}}, e_{\mathrm{in}})$ at three times. \emph{Bottom panel:} The orbital evolution of both planets over time.  
Compared to the dynamical mode shown in Figure \ref{fig:stability}, the more substantial early planet–planet interactions imprint themselves on the eccentricity of the inner planet. }
\label{fig:stability2}
\end{figure*}

\subsection{Individual Simulation Outcomes}
\label{sec:outcomes}
Of our 450 simulations, there are several distinct dynamical outcomes, resulting in systems that would be observed as two-planet systems (stable outcomes) and outcomes in which the inner planet is lost and the system would be observed as a lonely hot Jupiter (unstable outcomes). 

\subsubsection{Stable Dynamical Outcomes}
\label{sec:stable}

We define a stable evolution as one where both simulated planets remain bound in the planetary system, with the inner companion on a stable orbit interior to the Jupiter-mass planet (even if has a different final orbital radius or eccentricity from the initial system state).

\textit{Well separated, dynamically stable tidal evolution}. An example of this type of of interaction is shown in Figure \ref{fig:stability}. 
In this mode of evolution, there is sufficient separation between the orbital distance of the hot Jupiter and the orbital distance of the inner companion planet. While the planets do interact during the early stages of the Jupiter’s migration (evidenced by the characteristic libration in $\Delta\omega = \omega_{\mathrm{in}} - \omega_{\mathrm{out}}$ seen at early times), the inner planet does not attain a substantial orbital eccentricity during this process.

\textit{Highly interacting, dynamically stable tidal evolution}. 
In this dynamical mode, the system begins with planets more closely spaced than in the previous case. As a result, early planet–planet interactions lead the inner companion to attain a moderate ($e\sim 0.05 - 0.3$) to substantial ($e> 0.3$) orbital eccentricity. An example of this type of evolution is shown in Figure \ref{fig:stability2}, where early interactions drive the inner companion to a substantial orbital eccentricity (cycles with eccentricity maxima above $e\sim0.4-0.5$) some of which is retained even at later times, as the dominant behavior transitions into stellar-induced libration in $\Delta\omega = \omega_{\mathrm{in}} - \omega_{\mathrm{out}}$.

\subsubsection{Unstable Dynamical Outcomes}
\begin{figure*}[t!]
\centering

\includegraphics[width=0.97\textwidth]{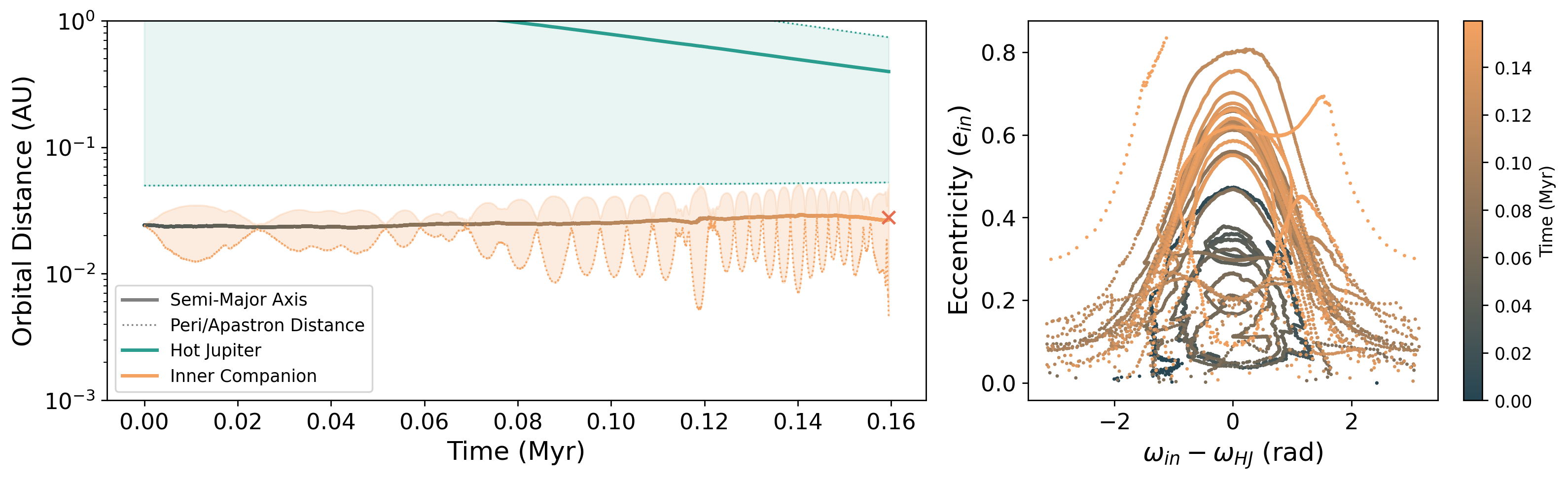}\\

\includegraphics[width=0.97\textwidth]{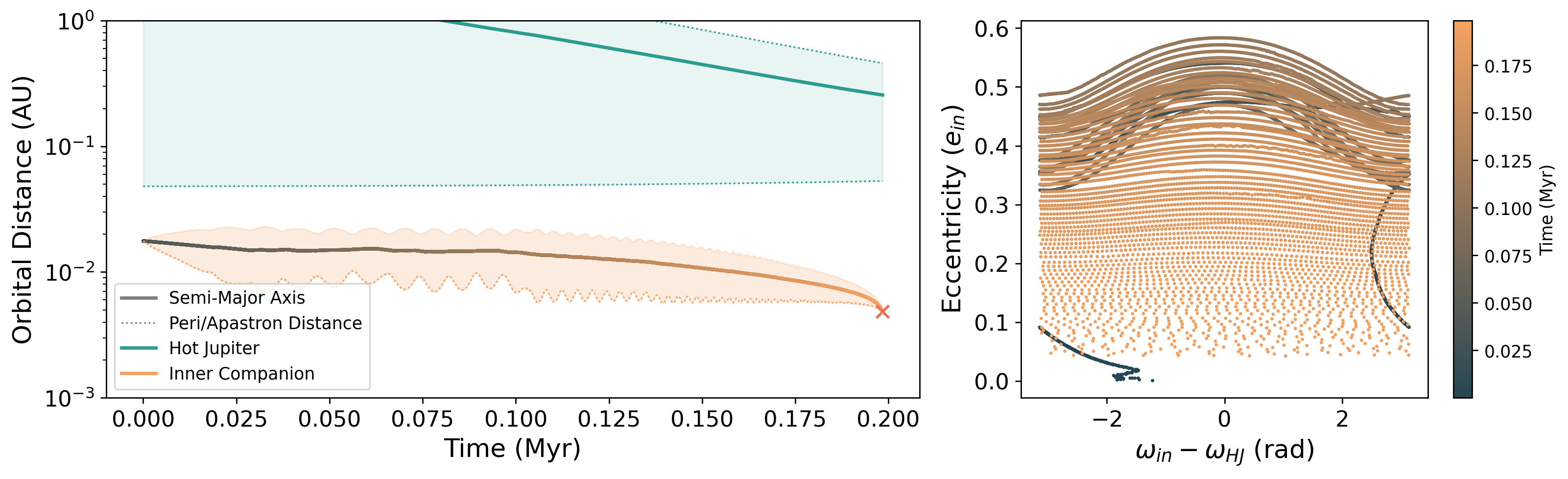}

\caption{Two examples of unstable evolutionary outcomes from our simulations. 
\textit{Top:} A system that becomes unstable due to dynamical perturbations, which excite large eccentricities and ultimately lead to the loss of the inner planet through direct collision with the central body. 
\textit{Bottom:} A system that becomes unstable primarily through tidal evolution, where the inner planet attains enough orbital eccentricity that tidal interactions dominate and it gradually spirals into the star. 
For both cases, the \textit{left} panels show the time evolution of the semi-major axis and peri/apastron distances of both planets, and the \textit{right} panels show the phase-space trajectory of the inner companion in the plane of eccentricity versus the difference in argument of periapsis relative to the hot Jupiter, color-coded by time. }
\label{fig:instability}
\end{figure*}

We consider a dynamical outcome as an ``unstable configuration" if the inner companion is either ejected from the system, collides with the hot Jupiter, or collides with the central body. 
This outcome occurs when the planets are too closely spaced initially, which leads to significantly perturbed orbits (more than the case shown in Figure \ref{fig:stability2}).
The two main dynamically unstable configuration sare shown in Figure \ref{fig:instability}: in the top panels, an integration demonstrating a case where the system went unstable due to the tidal evolution of the inner planet slowly onto a star-crossing orbit; in the bottom panels, direct planet-planet interactions led to the inner planet colliding with the central star. 

At early times, the phase space diagram shown in the top right panel of Figure \ref{fig:instability} exhibits libration in $\Delta\omega$-space, characteristic of secular interactions. 
At later times, the dynamics transition to circulation in $\Delta\omega$-space, a behavior more consistent with dynamics primarily driven by stellar effects (GR and tidal forces) but still influenced by planet-planet interactions. 
Compared to the smooth, regular contours seen in the stable cases (Figure \ref{fig:stability} and Figure \ref{fig:stability2}), the phase space in the unstable due to planet-planet interactions scenario (top panels of Figure \ref{fig:instability}) is irregular, a result of the disruptive influence of closer approaches between planets and subsequent stronger planet-planet interactions.
The bottom panels of Figure \ref{fig:instability} shows a simulation case where the inner planet is excited to a high eccentricity, then tidal interactions with the central star lead it to be engulfed. 

We note that it is also feasible that the hot Jupiter could be ejected from the system as well, but we do not see that outcome in our simulations. If the mass of our inner simulated planet were larger, that outcome might have been more likely. 

\subsection{Global Simulation Results}
In Figure \ref{fig:mean_stability}, we show the result of the entire parameter sweep. 
For visualization purposes, we bin the 450 simulations in $a_{\rm{in}},\ q_{\rm{out}}$ space. Within each bin, we compute the probability that a set of planets with the ascribed parameters would survive the tidal migration process. 
For each simulation, we categorize the final result as a ``survival" if the system has one the of the stable dynamical outcomes listed in Section \ref{sec:stable}. The remainder of the simulations are considered ``unstable configurations," and occur if the inner companion is either ejected from the system, collides with the hot Jupiter, or collides with the central body. 
Within each bin, we calculate the probability of survival by dividing the number of simulations where the configuration survived tidal migration by the total number of simulations in that bin.

\section{Case Studies: Candidates from the Observational Sample}
\label{sec:casestudies}
Having demonstrated in the previous section the feasibility of inner companions surviving the tidal migration of an outer Jupiter in general terms and for some parameter combinations, we next examine the observational sample to identify any candidate systems whose present-day architectures are consistent with assembly via tidal migration and the survival of an interior companion.
We conduct an initial analytic test to identify candidate systems that could have formed via this proposed mechanism. We begin by selecting systems from the exoplanet sample that contain a hot/warm Jupiter with a single known interior companion. We exclude systems with exterior adjacent companions, such as WASP-47 \citep{Becker2015}, since explaining the formation and migration of such additional planets would require further assumptions beyond the scope of this work.

In Figure \ref{fig:hill_spacing}, we show the mutual Hill radius separation between the inner companion and the hot/warm Jupiter at periastron, assuming the hot Jupiter's present-day semi-major axis was set by tidal migration. As a first-order filter, we assume that tidal migration is excluded in systems where the mutual Hill separation at the hot/warm Jupiter's periastron falls below the stability criterion of \citet{Chambers1996}. We also exclude systems where tidal migration would have caused orbit crossing, which are indicated by a negative initial $\Delta H$ in Figure \ref{fig:hill_spacing}.

After applying these criteria, we are left with three candidate systems: WASP-84, K2-290, and Kepler-418. We note that the results shown in Figure \ref{fig:mean_stability} suggest that none of these systems would have allowed an inner companion to survive the tidal migration of its outer companion, as all of these three systems have $\Delta_{H}^{\rm peri} < 14$. However, this figure was constructed with only one set of physical planet parameters. For completeness, in the following subsections we assess the feasibility of these individual systems having assembled via tidal migration.

\subsection{WASP-84}
The WASP-84 system contains two confirmed planets: a hot Jupiter, WASP-84b, with a current-day semi-major axis of 0.078~AU, radius of 0.96~$R_{J}$, and mass of 0.69~$M_{J}$; and an inner super-Earth, WASP-84c, with a semi-major axis of 0.024~AU, mass of $15\, \mathrm{ M}_{{\oplus }}$, and radius of $2\, \mathrm{ R}_{{\oplus }}$ \citep{Anderson2014, Maciejewski2023}.
The hot Jupiter was found to have a well-aligned orbit as compared to the spin axis of the star \citep{Anderson2015}, evidence that the system may have assembled via disk migration. 

With $a_{\mathrm{in}} = 0.024$~AU and an inferred past periastron distance of the hot Jupiter, $q_{\mathrm{HJ}} \approx 0.0389$~AU (assuming tidal migration occurred), WASP-84 falls within the region of parameter space in Figure~\ref{fig:mean_stability} where no analogous systems remain dynamically stable, since $\Delta_H^{peri} > 10$ at the inferred periastron distance. 

To further evaluate this, we ran ten integrations with the exact parameters of the WASP-84 system, starting the outer giant with $a_{\rm{out}} = 2$ AU and $q_{\rm{out}} = 0.039$ AU (the expected initial condition for a planet that post tidal migration would have the observed $a_{\rm{out}}=$0.078~AU). 
In all ten simulations, the inner planet ultimately became dynamically unstable due to interactions with the outer companion. This instability caused the inner planet’s orbit to attain significant eccentricity, which then allowed tidal evolution of the inner planet's orbit. This tidal evolution caused a rapid decay in semi-major axis, resulting in its eventual destruction in all ten integrations. 

To verify that this outcome was specifically driven by planet-planet interactions with the migrating Jupiter, we conducted an additional set of ten integrations in which the outer planet was not permitted to undergo tidal migration ($e_{\rm{out}}$ was set to 0). In these cases, the inner planet remained stable in its original orbit throughout the simulations. This suggests that in our previous set of simulations, the inward migration and forced eccentricity induced by the outer giant were responsible for the inner planet’s destruction via tidal evolution.
This result supports the conclusion of \citet{Anderson2015}, who argued that tidal migration could not have produced the observed configuration of the WASP-84 system.

\subsection{K2-290}
The K2-290 system contains two confirmed planets \citep{Hjorth2019}: an outer warm Jupiter, K2-290 c, with a semi-major axis of approximately 0.30~AU, a radius of 1.006~$R_{J}$, and a mass of 0.774~$M_{J}$; and an inner mini-Neptune, K2-290 b, with a semi-major axis of approximately 0.09~AU and a radius of 3.06~$R_{\oplus}$ \citep{Hjorth2019}. The mass of K2-290 b has not yet been directly measured.

Assuming the outer planet underwent tidal migration, we estimate a past periastron distance of $q_{\rm{out}} \approx 0.15$~AU. With $a_{\rm{in}} \approx 0.09$~AU, were the K2-290 system to have migrated via tidal evolution, it would have had a $\Delta_{H}^{peri} = 7.5$. Based on the boundary established in Figure~\ref{fig:mean_stability}, the system resides in a location in parameter space where stable co-existence of the inner planet and a tidally migrating outer companion is not observed. This suggests that K2-290 c likely did not undergo high-eccentricity tidal migration, and instead formed in situ or migrated through the protoplanetary disc.

To confirm this prediction, as we did for the WASP-84 system, we ran a suite of 10 simulations of the K2-290 system, using the measured parameter values except for an initial semi-major axis for K2-290 c of $a_{\rm{out}}=2$ AU and $e_{\rm{out}}=0.925$, to match its expected periastron distance ($q_{\rm{out}} \approx 0.15$~AU) were it to have migrated via tidal migration. None of the ten simulations allowed the inner companion to survive the tidal migration of K2-290 c.

Additionally, \citet{Hjorth2021} measured the stellar obliquity of K2-290 and found that the two transiting planets are in retrograde orbits (while still being aligned with each other), suggesting that the misalignment occurred coherently in the disk phase. 
This previous evidence and our simulations suggest that the K2-290 system also did not assemble via tidal migration.

\subsection{Kepler-418}
The Kepler-418 system contains two confirmed planets: an outer warm giant planet, Kepler-418 b, with a semi-major axis of 0.381~AU, radius of 1.20~$R_{J}$, and mass of 1.1~$M_{J}$; and an inner sub-Saturn, Kepler-418 c, with a semi-major axis of 0.102~AU and radius of 0.47~$R_{J}$ \citep{Tingley2014}. The mass of Kepler-418 c has not yet been measured, but for the purposes of our analysis, we assume that the planet is 0.5~$M_{J}$.

\begin{figure}[ht!]
\includegraphics[width=0.48\textwidth]{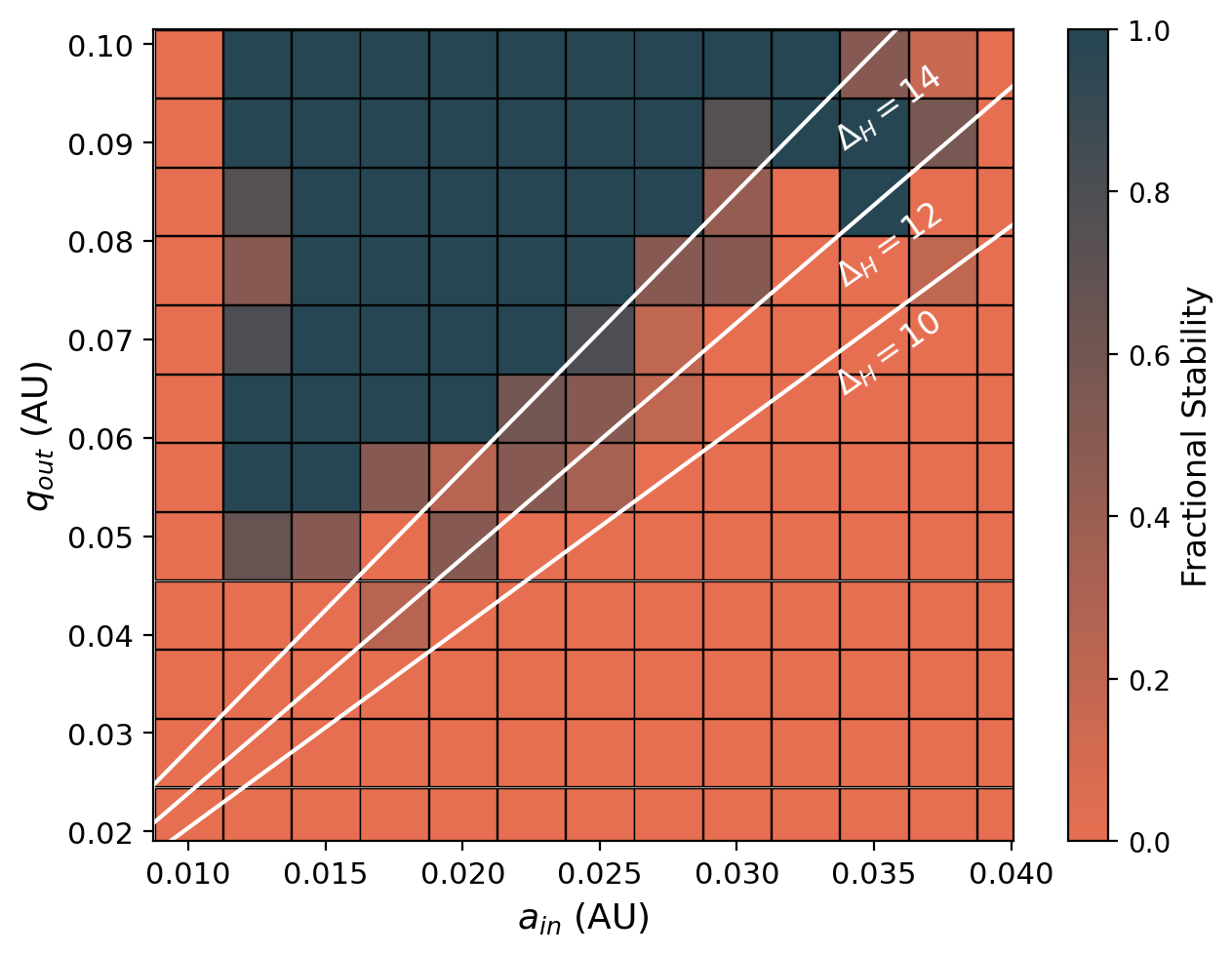}
\caption{We present the results of a parameter sweep study based on 450 numerical simulations, in which we vary the semi-major axis of the inner planet $a_\text{in}$ and the initial periastron distance of the hot Jupiter $q_\text{out}$. The simulations are binned and the fractional stability of the inner planet within each bin is denoted by the color scale. For comparison, we include the $\Delta_H = 10, 12, 14$ lines. Below the $\Delta_H = 10$ line, none of the simulations preserve the inner companion. However, above this threshold, there exists a region of parameter space where the survival of both planets is possible. Simulations above $\Delta_H = 14$ are likely to be stable, so long as $q_\text{out} \gtrsim\ 0.05$ The unstable region of the plot for all values of $q_\text{out}$ up to $a_{\rm{in}} \approx 0.01$ AU corresponds to initial conditions where the inner companion is unstable to tidal inspiral.  }
\label{fig:mean_stability}
\end{figure}

Assuming the outer planet underwent tidal migration, we estimate a past periastron distance of $q_{\mathrm{HJ}} \approx 0.1$~AU. With $a_{\mathrm{in}} = 0.102$~AU, the Kepler-418 system lies within the region of parameter space in Figure~\ref{fig:mean_stability} where stable co-existence of the inner planet and a tidally migrating outer companion is not generally observed (as $\Delta_H < 10$). We ran an additional ten simulations with the best-fit parameters described above, and none of the simulations allowed the inner planet to survive. Additionally, the circularization time for warm Jupiter Kepler-418 b would be expected to be somewhere around 10 Gyr (depending on its exact interior structure), making it unlikely that the system would be observed today in their co-transiting, circular orbits. For these reasons, it is also unlikely that Kepler-418 migrated via tidal high-eccentricity migration.

\section{Discussion}
\label{sec:discuss}
Our results emphasize a limitation (but not a very large one) to the assumption that high–eccentricity tidal
migration inexorably ejects or destroys interior planets.  
Both analytic stability criteria and direct \(N\)-body integrations show that inner companions can survive the tidal migration of an outer companion. However, the survival of such inner companions is only under rather specific circumstances.
Systems where tidal migration would imply that \(\Delta_{H}^{\rm peri}<10\) are unlikely to have formed via tidal migration, while those with \(\Delta_{H}^{\rm peri}> 14\) can survive if the migrating planet is a warm Jupiter (where $q_{\rm{out}} > 0.05$ AU). 

The absence of close inner companions around hot Jupiters is a natural consequence of high‐eccentricity tidal evolution, due to the small region of parameter space where companions may satisfy the \(\Delta_{H}^{\rm peri}> 14\) criterion along with the fact that much of that survival parameter space overlaps with the location where the inner planets themselves may tidally decay into the star. However, warm Jupiters enter the tidal‐migration phase with comparatively large periastra, potentially meeting this condition. However, they will be expected to circularize on $1-10+$ Gyr timescales and as a result will likely often retain measurable eccentricities (\(e\sim0.1\text{--}0.3\)) at the present day. 

Nonetheless, a narrow but non‐empty region of parameter space remains viable. Figure~\ref{fig:mean_stability} maps this region in the parameter space of initial periastron distance and inner‐planet semi‐major axis: warm/hot Jupiters beginning at \(\gtrsim\ 14\ \Delta_{H}^{\rm peri}\) and with periastron distances $0.5 < q_{\rm{out}} < 0.07$ will both allow an inner companion to survive, but also can finish circularizing within \(\lesssim1\)~Gyr or so. We note that these limits are approximate, depending on the exact tidal and orbital parameters in each system. Interior to this region, however, secular eccentricity pumping drives orbit crossings or direct stellar collisions. Exterior to this region, the warm Jupiter takes too long to circularize and would be seen as eccentric in the current day.


\subsection{Observational Prospects} 
In this work, we also examined the population of known warm- and hot-Jupiter systems with inner companions to determine if any of them could have been formed via tidal migration. 
Within our vetted sample of multi‐planet systems, three architectures were not immediately inconsistent with a high‐eccentricity pathway (WASP-84, K2-290, Kepler-418). However, our numerical simulations demonstrated that none of these systems likely formed in this way, as all three systems would have destabilized their inner planets during tidal migration.  Their present configurations thus \edit{likely} require alternative formation channels (e.g., in‐situ assembly or disk migration).
\edit{We will note that the individual simulations for each of these three systems may have had slightly different effective initial conditions than those implied by our analytic test in Section~\ref{sec:analytic}, as our mapping from the present-day semi-major axis of the giant planet to its assumed initial periastron distance assumes \(a_{\rm f} \simeq 2 q_{\rm init}\) and neglects angular-momentum transfer into the stellar spin. 
In the full tidal integrations, a fraction of the orbital angular momentum is instead deposited into stellar rotation, so that the same observed \(a_{\rm f}\) could in principle be reached from somewhat larger \(q_{\rm init}\). 
As a result, the system-specific integrations in this section tend to probe configurations that are slightly more compact than those expected from a fully self-consistent tidal reconstruction. However,  previous circumstantial evidence from the literature provides other lines of evidence supporting that these particular systems (WASP-84, K2-290, Kepler-418) did not form from tidal migration. }

Ultra-short period planets reside in 0.5\% of planetary systems orbiting G-dwarfs \citep{Sanchis-Ojeda2014}. Hot Jupiters are similarly rare \citep[with an occurrence rate somewhere around 0.3\% - 1\%;][]{Wright2012, Zhou2019, Beleznay2022}, while warm Jupiters are a few times more common \citep{Huang2016}. 
The covariance of those rates are not known, but the rarity of both types of objects are not working in our favor in terms of finding candidate systems that might seen the hot/warm Jupiter migrate tidally. Additionally, USP planets appear in general to be older than their nearby counterparts, suggesting that the standard USP-formation pathway operates over a wide range of stellar ages \citep{Tu2025}, which further reduces the population available at young ages when Jupiter might be tidally migrating to their final orbits. 

As a result, it is likely very uncommon that a planetary system with an interior short-period planet and outer hot- or warm-Jupiter assembles via tidal migration. Nonetheless, such configurations remain dynamically possible according to the results of our simulations, and future observations may identify candidates that could have formed through this pathway.

\subsection{Implications on Formation: Hot vs. Warm Jupiters}
Keeping multi-planet systems containing Jupiters stable is a real problem, even for systems that migrate in disk-driven ways. Similarly to our simulations that model the tidal evolution post-disk-dispersal, the simulations of \citet{He2024} modeling the disk phase found that multi-planet systems ending as warm Jupiters more readily survive compared to those where the migrating Jupiter ends in a shorter-period, hot orbit. This is analagous to the results of our simulations (which are of a later evolutionary stage of the system), where warm Jupiter systems more readily allow inner companions to survive via tidal migration. 

\citet{Petrovich2016} found that, based on parameter distributions, only warm Jupiters with the 20\% highest eccentricities likely formed via tidal migration, and the remainder formed via a different mechanism.
In this work, we find that the parameter space for inner companions to survive tidal migration is small, and the ones that do survive generally have companion Jupiters with longer circularization timescales. Since $\tau_{circ} \propto a^{-13/2}$, the warm Jupiters will take substantially longer to circularize than the hot Jupiters. As a result, while inner companions can survive in these systems, it is likely, as \citet{Petrovich2016} found, that the warm Jupiters that do survive will have relatively substantial eccentricities today. 

Recent similar work by \citet{Zhu2025}\footnote{completed simultaneously with the preparation of this manuscript} found that with a very similar simulation geometry to that considered in this work, the tidal migration of hot Jupiters with inner companions allows the survival of the inner companions as USP planets. \citet{Zhu2025} considered much lower initial eccentricities ($e_{\rm{out}}\sim 0.05 - 0.2$) for the migrating Jupiter than those considered in this work ($e_{\rm{out}}\sim 0.95 - 0.99$).
Our results show that survival is also possible for warm Jupiters (including those very close to the dividing line between warm and hot Jupiters) even with the much higher initial eccentricities characteristic of an origin closer to the ice line. 

\subsection{Future Work}
In this study, we focused on coplanar configurations, which are supported by observations suggesting that many hot Jupiters that undergo tidal migration do so in a coplanar manner \citep{Zink2023}. 
Future work should explore the impact of non-coplanar architectures on the survival of inner companions during high-eccentricity migration. Inclined orbits are expected to exhibit different instability dynamics than the coplanar cases considered here \citep{Ito1999}. A gas giant on an inclined orbit may also induce Kozai–Lidov cycles \citep{Kozai1962, Lidov1962}, which could either further destabilize the inner planet or, through additional dynamical pathways available in three dimensions, promote long-term stability.

\edit{In addition, our current model assumes a constant tidal quality factor \( Q \) and does not explicitly account for the evolution of the stellar spin. In reality, both \( Q \) and the stellar rotation rate evolve as the orbit circularizes, which can modify the migration and synchronization timescales \citep[e.g.,][]{Naoz2012}. Although we rescaled all simulations to enforce circularization within a fixed 1~Myr integration window (thereby removing our sensitivity to the absolute circularization time) future work should relax this simplification, as the dynamics may change depending on the rate at which the circularization occurs.}

\edit{Our strategy of rescaling the integration lengths, necessary due to computational limitations to run the large parameter sweep presented in Figure \ref{fig:mean_stability}, has the potential to underestimate the stability of some of the shortest-period and most quickly evolving planets. This means that while the regions of stability presented in Figure \ref{fig:mean_stability} are secure, it is possible that some system orientations in the `unstable' regions of parameter space could actually survive for some system configurations.}

\edit{Finally, in this work, our main simulation suite considers a fiducial case of the Sun-like host star with a circularizing inner Earth-like planet and out Jupiter-like planet. In reality, the exoplanet census is much more diverse than this, and planet masses and radii vary substantially. Future work should determine how Figure \ref{fig:mean_stability} changes with varying physical planet parameters. }

\section{Conclusion}
\label{sec:conclude}

In this work, we consider whether inner inner planets may survive the tidal migration of their outer, Jupiter-mass companions.
Our derived analytic criterion and suite of N-body simulations find conditions under which such systems may survive. High-eccentricity tidal migration is invariably fatal to interior companion planets if the  $\Delta_{H}^{\rm peri}\!\approx\!10$, but systems with $\Delta_{H}^{\rm peri}\!>14$ and $q_{\rm{out}} \gtrsim  0.05$ may allow both components to survive. 
All twelve observed systems that host both a short-period giant and a confirmed interior planet fail this criterion, and follow-up integrations demonstrate that none of their current architectures could not have emerged from high-eccentricity tidal evolution.  
More broadly, our results provide an easy, analytical test (comparing the computed $\Delta_{H}^{\rm peri}\!$ against the conditions for total instability and survival) that distinguishes systems that must have migrated via some disk-dominated mechanism from those who could have assembled via tidal migration.

\begin{acknowledgments}
J.B. would like to thank Michael Dieterle, Konstantin Batygin, Fred Adams, Thomas MacLean, and Devansh Mathur for useful conversations.
\end{acknowledgments}

\begin{contribution}
All authors contributed equally to this work.

\end{contribution}

%

\facilities{Exoplanet Archive \citep{Christiansen2025}}

\software{
\texttt{matplotlib} \citep{Hunter:2007},
\texttt{pandas} \citep{mckinney-proc-scipy-2010, the_pandas_development_team_2024_13819579},
\texttt{REBOUND} \citep{Rein_2012}
\texttt{REBOUNDx} \citep{Tamayo2020}, \texttt{scipy} \citep{2020SciPy-NMeth}}


\bibliography{sample7}{}
\bibliographystyle{aasjournalv7}



\end{document}